# Hard Rock Drilling for Super-hot Enhanced Geothermal System Development: Literature Review and Techno-Economic Analysis

**Orkhan Khankishiyev, Saeed Salehi**

**Well Construction Technology Center - The University of Oklahoma, Norman, OK, USA**

## Keywords

## ABSTRACT

The increasing global demand for electricity and the imperative of achieving sustainable and net-zero energy solutions have underscored the importance of exploring alternative sources. Enhanced Geothermal Systems (EGS) have emerged as a promising avenue for renewable and sustainable energy production. However, the development of EGS faces a significant challenge in drilling through hard rock formations at high temperatures, necessitating specialized drilling equipment and techniques.

This study aims to investigate the current state-of-the-art technology for drilling in hard rock formations under elevated temperatures, specifically in the context of super-hot EGS development. It involves a comprehensive review of previous projects and a meticulous analysis of existing drilling technologies and techniques. Furthermore, a techno-economic evaluation will be conducted to assess the feasibility of super-hot EGS development in hard igneous formations, considering key factors such as drilling performance, operational challenges, and material costs.

The outcomes of this study will enhance the understanding of the technical challenges associated with super-hot EGS development and facilitate the design of efficient and cost-effective drilling technologies for the geothermal energy industry. By improving the drilling process in EGS development, the full potential of geothermal energy can be harnessed as a viable and sustainable energy source to meet the growing global demand for electricity.

# 1. Introduction

With a rising number of population and immensely growing global economy, the demand for sustainable energy is increasing. The theory and calculations on paper indicates that the planet Earth has an enormous amount of sustainable and recoverable geothermal energy that is capable of meeting the heat and electricity demand of humanity for a long time (Tester et al., 2006). Special emphasis should be placed on the term "recoverable" which heavily relies on the existence of sophisticated technology because there are plenty of wells that are producing geothermal energy but the enthalpy recovery efficiency decreases over time (Soltani et al., 2019) due to limited well depth and technology. The existing geothermal power plants use the conventional Rankine cycle technology to transform the geothermal energy of the resources with temperatures between 125°C and 200°C into electricity. With low energy density of these temperatures, the plant efficiency changes from 8% to 15% net thermal efficiency ($\eta_{th}$), and it makes the capital cost of power very high. By tapping into resources that have an order of magnitude more energy potential than conventional systems, super-hot enhanced geothermal systems will be economically efficient with capability to provide cheapest (46 $/MW-hr.) electricity to the end-users (AltaRock Energy, 2017).

On the other hand, remarkably bigger geothermal potential can be untapped by drilling deeper to 500°C temperature. Several countries around the world are currently racing to investigate the potential of deep and super-hot drilling and prove the feasibility of the concept. United States, Japan, Iceland, Mexico, Italy and New Zealand are some of the countries that are dedicated to pioneering the technology for the super-hot Enhanced Geothermal System development in sedimentary and igneous basins with temperatures above 400°C (Petty (a) et al., 2020).

The abovementioned temperatures can be encountered at shallower depths when they are associated with volcanic magma intrusions that threaten the safe drilling operation. The IDDP-1 well was forced to be abandoned despite of the huge economic investment when the drilling hit the magma intrusion and well started to produce corrosive super-critical steam that destroyed the well (Friðleifsson et al., 2015). The amount of geothermal energy stored in igneous basins represents the big portion of the recoverable enthalpy but there is significantly less drilling practice and more risk for the drilling operations in this type of rocks. On the other hand, deep layers of the sedimentary basins (>10 km) hold temperatures above 375-400°C that are suitable for super-hot EGS (Blackwell et al., 2011). There are decades of drilling experience in sedimentary rocks while only handful of the High Pressure High Temperature (HPHT) wells were drilled (Madu & Akinfolarin, 2013) leaving the super-critical temperature reservoirs unexplored.

Achieving successful drilling operations in super-hot sedimentary and igneous basins requires the availability of robust drilling technologies capable of withstanding harsh conditions. Essential technologies include drill bits, drill strings, drilling fluids with consistent properties, directional drilling tools for profile creation, and logging and measurement tools for surveying well deviation and formations drilled. Moreover, advancements in resource characterization, near-term and long-term field development, reservoir development and management, and efficient energy conversion are imperative for the realization of super-hot enhanced geothermal power production. Additionally, continuous research and development in areas such as geomechanics and drilling optimization techniques are crucial to ensure efficient drilling operations and maximize the potential of super-hot enhanced geothermal systems.

This study aims to investigate the state-of-the-art technology in hard rock drilling under elevated temperatures for the development of super-hot enhanced geothermal systems. Through a comprehensive review of previous projects and a detailed analysis of existing drilling technologies and techniques, valuable insights into the technical challenges associated with super-hot enhanced geothermal system development will be gained. Additionally, a techno-economic evaluation will be conducted to assess the viability of super-hot enhanced geothermal systems in hard igneous formations, considering drilling performance, operational challenges, and material costs. The findings of this study will contribute to the design of efficient and cost-effective drilling technologies, unlocking the full potential of geothermal energy as a renewable and sustainable energy source to meet the growing global demand for electricity.

## 2. Geothermal Energy and Hard Rock Formations

### 2.1. Overview of geothermal energy and its potential

Geothermal energy represents a significant source of renewable energy that harnesses the heat stored within the Earth's crust. The heat originates from the radioactive decay of elements such as uranium, thorium, and potassium, as well as residual heat from the planet's formation. Geothermal resources can be found globally, and their utilization offers a reliable and sustainable alternative to conventional fossil fuels. The potential of geothermal energy is vast, with estimates suggesting that the heat content of the Earth's uppermost six kilometers is equivalent to 50,000 times the energy stored in all known oil and gas reserves (Tester et al., 2006). Geothermal resources are categorized into three main types: hydrothermal systems, enhanced geothermal systems (EGS), and deep geothermal systems. Hydrothermal systems are the most common and accessible type, consisting of naturally occurring reservoirs of hot water or steam. These resources are typically found in areas with active tectonic activity, such as geothermal fields and volcanic regions. Hydrothermal systems have been successfully utilized for power generation in various countries, with installed capacity totaling several gigawatts worldwide.

Enhanced geothermal systems (EGS) offer the potential to access deeper and hotter resources that are not naturally present in hydrothermal systems. EGS involves creating an artificial geothermal reservoir by injecting fluids into hot and permeable rock formations, stimulating the flow of heat to production wells (Lu, 2018). This technology allows for geothermal energy extraction in areas where conventional hydrothermal resources are limited. Deep geothermal systems, also known as hot dry rock systems, involve extracting heat from impermeable rocks by creating a reservoir through hydraulic fracturing and injecting fluid to extract heat. This approach enables the utilization of geothermal energy in areas with low natural permeability (McClure & Horne, 2014).

The development and utilization of geothermal resources depend on a comprehensive understanding of the subsurface conditions, including rock properties, fluid characteristics, and the presence of geological structures (Lund et al., 2008). Geological surveys, seismic studies, and other exploration techniques are employed to assess the potential of a given area for geothermal energy production. In recent years, there has been growing interest in the development of super-hot enhanced geothermal systems (SH-EGS) in hard rock formations. These systems aim to access higher temperature resources (exceeding 400°C) that offer greater energy potential and improved efficiency in power generation (Kumari & Ranjith, 2019). However, drilling and operating in such extreme conditions present significant technical challenges that require advanced drilling technologies and techniques.

## 2.2. Hard rock formations as suitable reservoirs for super-hot geothermal system

Hard rock formations offer promising potential as suitable reservoirs for super-hot geothermal systems due to their unique characteristics and thermal properties. These formations, typically composed of igneous or metamorphic rocks, exhibit high temperature gradients and enhanced thermal conductivity, enabling the extraction of geothermal heat at elevated temperatures (Feng et al., 2022). Igneous rock formations, such as granite and basalt, possess excellent thermal properties that make them ideal candidates for super-hot geothermal reservoirs. These rocks have high heat capacity and thermal conductivity, allowing for efficient heat transfer from the surrounding hot rocks to the produced fluid (Sipio et al., 2013; Sundberg et al., 2009). The ability of igneous rocks to sustain high temperatures over long periods of time makes them attractive for super-hot geothermal energy production. The suitability of hard rock formations as reservoirs for super-hot geothermal systems is further enhanced by their geological stability and durability. Compared to sedimentary formations, which may exhibit structural instability and compaction, hard rocks provide a more reliable and long-lasting reservoir for sustained geothermal operations (Das & Chatterjee, 2017; Ma et al., 2022). The structural integrity of hard rock formations minimizes the risk of well collapse and maintains the permeability necessary for fluid flow.

Furthermore, the potential of hard rock formations for super-hot geothermal systems is closely linked to the presence of natural heat sources, such as magmatic intrusions. These intrusions create zones of elevated temperatures within the rocks, enabling the development of high-enthalpy reservoirs (Friðleifsson et al., 2015). However, drilling in the vicinity of magmatic intrusions poses challenges, as the extreme heat and corrosive nature of supercritical fluids can pose risks to well integrity and equipment (Pálsson et al., 2014).

## 2.3. Challenges associated with developing super-hot enhanced geothermal systems

(Kruszewski & Wittig, 2018) conducted a comprehensive analysis of failure modes in 20 high-enthalpy geothermal wells worldwide, including locations in Iceland, Italy, and Japan, which experienced temperatures above the critical point. It is crucial to define the term "super-critical" accurately, as it is often used interchangeably with "super-hot." The critical point of pure water occurs at a temperature of 374°C and a pressure of 221 bar. However, the presence of salts can increase the critical temperature and pressure, resulting in even deeper drilling requirements for super-critical geothermal wells. For instance, (Bischoff & Rosenbauer, 1984) found that seawater with a NaCl concentration of 3.5% reached critical conditions at 405°C and 302 bar. The failure modes of geothermal wells are closely linked to the temperature and composition of the geothermal fluid/steam, which can damage drill bit/drill string components, drilling fluids, casing/cement, downhole production systems, and surface drilling and production systems. Common failure modes observed in previous projects include metal/elastomer fatigue, deterioration of drilling fluids, casing fatigue caused by temperature, cement bond failure due to temperature, significant corrosion of metal components due to the acidity of geothermal fluid, partial and complete loss circulation due to extensive fracture networks in igneous rocks, and scale accumulation in surface production equipment.

Figure 1 below illustrates the maximum reservoir temperatures and pressures recorded in geothermal wells where supercritical conditions were encountered. The blue line represents the critical point of clean water, while the red line represents the critical point of seawater. Given the limited number of geothermal wells drilled to supercritical conditions, there is a scarcity of industry standards or common practices for geothermal well design and drilling operations. Near the brittle-ductile transition zone, located a few kilometers below the Earth's crust, magmatically dominated fluids exist within hotter plastic rocks, while hydrothermal fluids flow through the underlying colder brittle rocks (Fournier, 1999). As a result, supercritical conditions develop in this region. There have been instances where unexpectedly encountering supercritical temperatures and/or pressures during drilling operations for exploration and production wells.

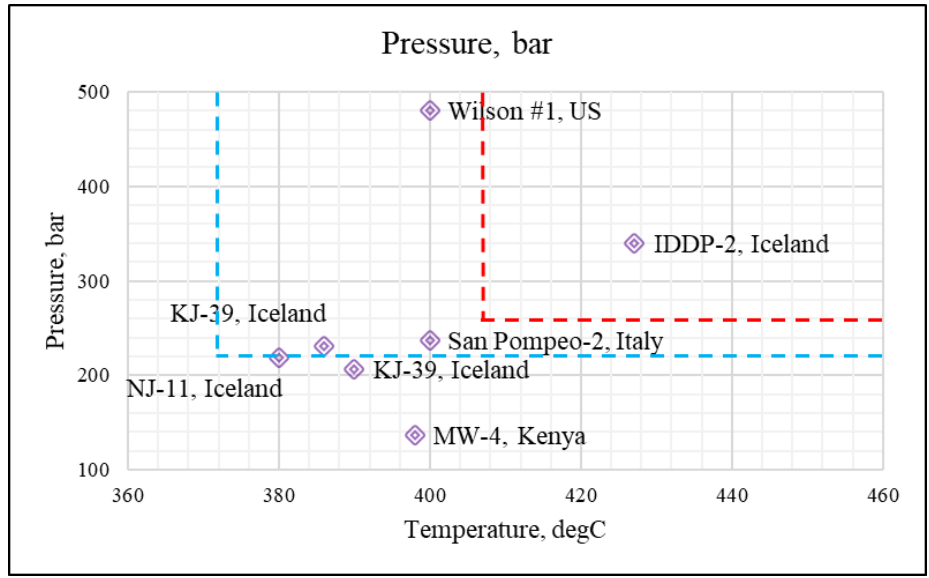

**Figure 1. Maximum reservoir temperatures and pressures measured in geothermal wells. (Kruszewski & Wittig, 2018)**

Challenges commonly experienced during these operations are often attributed to the physical properties of the rock and fluid, leading to failures in drilling, completion, or reservoir fluid handling. Some wells have reported dry conditions, indicating a lack of either sufficient permeability or formation fluid with adequate reservoir pressure, both of which are essential elements for efficient enhanced geothermal systems (EGS) applications.

### 2.3.1. Abrasive rock formations and temperatures

Super-hot geothermal systems often involve drilling through abrasive rock formations, which can accelerate drill bit wear and impact drilling efficiency. Additionally, the high temperatures can exacerbate the abrasive nature of the formations, further challenging drilling operations (Cardoe et al., 2021). Implementing robust drill bit designs, advanced drilling fluids with abrasion-resistant properties, and appropriate drilling parameters can help overcome abrasive rock challenges.

### 2.3.2. Drill string and Casing Material Failure

The extreme temperatures and demanding downhole conditions in super-hot geothermal systems can subject the drill string and casing materials to significant mechanical and thermal stresses, leading to material failure (Torres, 2014). Premature drill string failures and casing deformations can compromise drilling operations and wellbore integrity. Employing high-strength materials, proper design considerations, and thermal insulation techniques can enhance the reliability and performance of drill strings and casing materials.

### 2.3.4. Wellbore instability

Drilling in super-hot conditions can lead to wellbore instability, particularly in hard rock formations. Elevated temperatures and high-pressure environments can induce rock stress relaxation, causing wellbore collapse, formation damage, and reduced wellbore stability (Wu et al., 2020; Yan et al., 2014). Mitigation measures such as proper wellbore support, casing design, and drilling fluid selection are crucial to prevent wellbore instability and maintain wellbore stability.

### 2.3.4. Loss circulation events during drilling

Loss circulation refers to the unintended loss of drilling fluids into permeable formations, resulting in reduced drilling efficiency, lost circulation zones, and potential well control issues. In super-hot geothermal systems, loss circulation events can be particularly challenging due to the high temperatures and complex rock formations (Magzoub et al., 2021). Effective wellbore strengthening techniques, such as wellbore strengthening agents and lost circulation materials, can help mitigate loss circulation challenges.

### 2.3.5. Deterioration of Drilling Fluids

The harsh conditions and high temperatures in super-hot geothermal systems can cause the deterioration of drilling fluids. Factors such as thermal degradation, chemical reactions, and contamination can lead to a decrease in drilling fluid performance and impact drilling operations (Mohamed, Salehi, & Ahmed, 2021). Implementing appropriate drilling fluid selection, regular testing, and conditioning procedures are necessary to mitigate the deterioration of drilling fluids and ensure their effectiveness during drilling.

### 2.3.6. Wellbore Integrity: Cement Bond Failure

Maintaining proper wellbore integrity is crucial in super-hot geothermal systems, and cement bond failure can pose significant challenges. High temperatures and thermal cycling can lead to cement degradation, loss of bond strength, and potential fluid migration along the wellbore (Kang et al., 2022; Petty (b) et al., 2020; Shah, 2021). Employing appropriate cementing techniques, using thermally stable cement formulations, and implementing effective quality control measures are vital to prevent cement bond failure and ensure wellbore integrity.

### 2.3.7. Corrosion and erosion

Super-hot geothermal environments often involve exposure to corrosive and abrasive fluids. The presence of aggressive chemicals and high-velocity fluid flows can lead to corrosion and erosion of wellbore materials, including casing, drill bits, and downhole equipment (Karlsdóttir et al., 2019). Utilizing corrosion-resistant materials, implementing protective coatings, and optimizing fluid compositions can minimize corrosion and erosion effects.

### 2.3.8. Scaling and mineral deposition

Geothermal fluids often carry dissolved minerals that can precipitate and deposit on wellbore surfaces and within the reservoir, resulting in scaling and reduced permeability (Klapper et al., 2019). Scaling can impede fluid flow, decrease heat transfer efficiency, and lead to equipment fouling. Proper fluid chemistry management, scale inhibitors, and regular maintenance are essential for mitigating scaling issues.

### 2.3.9. Reservoir performance and sustainability

The long-term performance and sustainability of super-hot enhanced geothermal systems heavily rely on efficient reservoir management and production optimization. Challenges in this area include reservoir characterization, fluid extraction techniques, reservoir pressure maintenance, and long-term sustainability (Petty (b) et al., 2020). Implementing effective reservoir monitoring, suitable reinjection strategies, and advanced reservoir modeling can contribute to maximizing reservoir performance and long-term sustainability.

### 2.3.10. Seismicity and induced seismic events

Intensive geothermal operations, including drilling and reservoir stimulation, can induce seismic activity in some cases. The interaction between injected fluids, rock fractures, and pre-existing faults can trigger seismic events, ranging from microseismicity to larger magnitude earthquakes (Kim et al., 2018; Majer et al., 2007; Sherburn et al., 2014). Monitoring and characterization of seismicity, along with proper reservoir management strategies, are necessary to minimize the risk of induced seismic events.

## 3. Technology Review

### 3.1. Limitations of conventional geothermal systems

An overview of the conventional systems was provided in Section 2.1. Conventional geothermal systems have played a crucial role in harnessing geothermal energy, but they possess certain limitations that super-hot enhanced geothermal systems aim to overcome. One major limitation of conventional geothermal systems is their reliance on naturally occurring high-temperature resources, which are often geographically constrained to specific regions. This limitation restricts the widespread adoption of geothermal energy and hampers its potential as a global renewable energy source. In contrast, super-hot enhanced geothermal systems have the advantage of being able to utilize hard rock formations found in a broader range of geographical locations. This wider accessibility allows for the development of geothermal projects in areas that were previously considered unsuitable for conventional systems, thereby expanding the reach of geothermal energy and maximizing its utilization (Tester et al., 2006).

Another limitation of conventional geothermal systems is the decline in reservoir performance and energy output over time. Continuous extraction of geothermal fluids from conventional reservoirs can lead to pressure drawdown and decreased reservoir temperatures, resulting in reduced power generation capabilities. Super-hot enhanced geothermal systems address this limitation by targeting higher temperatures and more abundant heat resources in hard rock formations. These systems can tap into super-hot zones, where temperatures exceed those typically encountered in conventional geothermal reservoirs. By accessing higher-temperature resources, super-hot enhanced geothermal systems have the potential to enhance overall energy production and extend the lifespan of geothermal projects, making them more economically viable and sustainable in the long term (Cladouhos, 2017).

### 3.2. Summary of super-hot geothermal projects

| # | Project name / Country | Year | Well Names | Number of wells | Depth, ft | Temp., °F | Pres., psi | Lithology |
|---|---|---|---|---|---|---|---|---|
| 1 | Puna / Puna, Hawaii, U.S. | 1976-2005 | - | 8 | 6456-8389 | 649-1922 | 420-2300 | Fine, hypidiomorphic granular basalts dominated with crystals of plagioclase and pyroxene |
| 2 | The Geysers / Mayacamas Mtns, CA, U.S. | 2010 | Prati-32 (re-drill) | 1 | 11142 | 752 | N/A | Fractured metamorphosed basalt, shale, and greywacke |
| 3 | Salton Sea / Imperial Valley, CA, U.S. | 1964-1965, 1990 | Elmore #1, IID-14 | 2 | 6801-7119 | 680-734 | 3002-3089 | Plio-Pleistocene sediments, deltaic, and quartzofeldspathic sandstones, with clay or carbonate cements, interbedded with lacustrine mudstones and siltstones |
| 4 | Los Humeros / Los Humeros Basin, Mexico | 1985-2008 | - | 13 | 6289-9843 | 605-854 | 1639-3191 | Andesites, basaltic to dacitic lavas |
| 5 | Reykjanes / Reykjanes Peninsula, Iceland | 2005, 2016-2017 | RN-17, IDDP-2 | 2 | 10113-15285 | 698-799 | 4931 | Crystallized pillow basalts, sedimentary tuff formations |
| 6 | Krafla Magma Testbed / Krafla, Iceland | 1982-2008 | KMT-1, IDDP-1 | 5 | 6345-9377 | 669-1652 | 1798-3336 | Volcaniclastic breccias, basaltic lavas which are mostly olivine tholeiites |
| 7 | Larderello (DESCRAMBLE) / Larderello, Italy | 1980-2017 | Venelle-02, Sasso-22, San Pompeo-2 | 6 | 4199-15912 | 698-959 | 1247-4351 | Fractured quartzitic phyllites metamorphic basement, Granites were drilled occasionally |
| 8 | Nisyros Island / Nisyros, Greece | 1982 | Nisyros-1 | 1 | 5958 | 750 | N/A | Basaltic and andesitic pillow-lavas |
| 9 | Kenya / Menengai, Kenya | 2011 | MW-01, MW-03, MW-04, MW-06 | 4 | 6909-7205 | 662-752 | 2031 | Trachyte, major reservoir regions occur at the quartz-illite epidote zone |
| 10 | Kakkonda / Kakkonda, Japan | 1994-1997 | WD-1A, WD-1b (sidetrack) | 2 | 9721-12234 | 662-1076 | N/A | Quaternary granite intrusion |

**Table 1. Summary of super-hot geothermal projects: project and well names, number of wells, depth, temperature, pressure and lithology (Kruszewski & Wittig, 2018), (CATF, 2022)**

| # | Project name / Country | ROP, ft/day | Casing size | Drilling Fluid | Major Issues |
|---|---|---|---|---|---|
| 1 | Puna / Puna, Hawaii, U.S. | ~220 | 8 1/2 in open hole and/or 7 in perforated uncemented liner | Water, aerated mud, or foam at shallower high permeability zones, Water Based Mud for the remaining intervals | Loss circulation at shallower high permeability zones |
| 2 | The Geysers / Mayacamas Mtns, CA, U.S. | ~120-480 | 7 in slotted liner | Air at deeper high temp zones | Severe bit wearing during air drilling at high temp zones |
| 3 | Salton Sea / Imperial Valley, CA, U.S. | ~100-180 | 7 in liner + 6 1/8 in Open hole | LCM was used continuously. No info on drilling fluid type. | Failed well logging attempts due to mud density and temperatures above 752 °F. Drilled some intervals without circulation. Several failed coring attempts. |
| 4 | Los Humeros / Los Humeros Basin, Mexico | N/A | N/A | N/A | N/A |
| 5 | Reykjanes / Reykjanes Peninsula, Iceland | ~200-320 | 7 in slotted liner + 6 in open hole | Water Based Mud for shallower zones, water for severe loss circulation zones | Total loss circulation below 10500 ft. Frequent fluid losses due to naturally fractured formations. Inefficient core recovery |
| 6 | Krafla Magma Testbed / Krafla, Iceland | ~160-390 | 9 5/8 in slotted liner | After 6700 ft, the water-based drilling mud was replaced with water because of total loss circulation | Total loss circulation in heavily fractured zones. Drilling into magma chamber, tool stuck, and side-tracking three times. Failed core recovery attempts, heavy core bit wearing |
| 7 | Larderello (DESCRAMBLE) / Larderello, Italy | N/A | 7 in liner + 6 in open hole | Water Base Mud weighted up with Ilmenite (Microdense™) and with Sepiolite as suspending agent, replaced with water at high temp deeper zones | Stuck pipe for differential pressure, fluid loss, total loss circulation. LCM failed to block the fractures |
| 8 | Nisyros Island / Nisyros, Greece | N/A | 7 in slotted liner | Water based mud | Casing collapse and buckling due to fast heating and cooling |
| 9 | Kenya / Menengai, Kenya | 200-290 | 7 in slotted liner | Aerated water and foam for loss circulation zones, water-based mud for remaining intervals | Stuck pipe when drilled into magma chambers, partial and total loss circulations, drill bit damage due to high temperatures |
| 10 | Kakkonda / Kakkonda, Japan | 60-345 | 9 5/8 in casing + 8 1/2 in open hole | Water based mud | Mud deterioration, drilling was terminated at 12234 ft because of H2S gas influx |

**Table 2. Summary of super-hot geothermal projects: rate-of-penetration (ROP), drilling fluid types, and major drilling challenges, (Kruszewski & Wittig, 2018), (CATF, 2022)**

Tables 1 and 2 summarize the details of previous super-hot geothermal projects including the lithology, rate of penetration, drilling fluid and major challenges faced during drilling. The hardness of the rock together with temperature made the ROP values significantly smaller than what is usually encountered in conventional sedimentary rock drilling practices (Khankishiyev et al., 2023). Severe loss circulation accidents due to natural fractures were also common among super-hot drilling projects.

Loss of circulation is a prevalent and significant challenge in the drilling of super-hot geothermal wells due to the widespread natural and thermally-induced fracture networks present in basaltic igneous rocks. The cost of loss of circulation can exceed 20% of the total expenditure for exploration well drilling (Lavrov, 2016). An investigation by (Cole et al., 2017) revealed that natural fractures were the primary cause of loss circulation in geothermal wells drilled between 2009 and 2017. Despite attempts to block fractures using various Loss Circulation Materials (LCMs) or seal them through cementing, complete loss of circulation has been reported during the drilling of major super-hot geothermal wells such as Sasso 22 by (Batini et al., 1983), the Descramble project by (Bertani et al., 2018), IDDP-1 and IDDP-2 wells by (Friðleifsson et al., 2015) and (Friðleifsson et al., 2017). Therefore, the drilling operations at the loss circulation intervals were carried out using water to prevent the high cost of WBM and OBM.

Previously, the Sasso 22 and San Pompeo 2 wells were drilled to 4092 m and 2930 m in Italy with the purpose of exploring an EGS resource (Bertini et al., 1980). Downhole temperatures and pressures up to 394 °C and 212 bar were measured at 2560 m depth in the San Pompeo 2. The Sasso 22 well was drilled without circulation and due to elevated temperatures and corrosive conditions, severe drilling problems such as tool deviation, drill pipe corrosion, breakage, fishing, and side tracking arose below 3000 m. Violent blowout happened while drilling the San Pompeo 2 due to inhomogeneous heavily fractured rock and both wells were eventually abandoned (Batini et al., 1983).

The IDDP-1 well drilled to 2100 m in Iceland accidentally hit a magma intrusion of the volcano. The drilling stopped due to supercritical flow from the well. The well achieved flow rates of up to 50 kg/s at temperatures of up to 452°C. Sustained flow rates of ∼30 kg/s could generate ∼20 MWe of power while the strongly corrosive steam destroyed the well and it was abandoned (Friðleifsson et al., 2015). The IDDP-2 well was drilled and completed at 4659 m without circulation. The temperature of 427°C and pressure of 4930 psi was recorded at the bottomhole. The well was put on hold without carrying out any flow test (Friðleifsson et al., 2017).

Another example of extreme temperature drilling is WD-1 well drilled to the depth of 3729 m, 500°C in at the Kakkonda geothermal field in Japan as part of the Deep Geothermal Resources Survey led by NEDO (Muraoka et al., 2014). The bottom part of drilling operation took place below the brittle-ductile transition zone with lower fracture density by cooling down the bottom hole assembly (BHA) with the drilling fluid (Madu & Akinfolarin, 2013). Although the well did not produce any supercritical fluid, it was a great demonstration of the possibility of drilling below the brittle-ductile transition zone. The DESCRAMBLE project in Italy is a prime example of the hottest geothermal well ever successfully completed. The well was completed at 2900 m in rock that was over 514°C in temperature to prove the concept and test the latest technology in action. After being finished in 2017, Venelle-2 is suspended waiting for the next steps. With financing from the EU, the GEMex project is still active and aims to produce above supercritical temperature (Bertani et al., 2018).

## 3.3. Summary of corrosion findings

| Well Name / Location | MD, m/ft | Temp., °C/°F | Casing/Tubing Size, in | Casing/Tubing Material | Failure mode | Damage location | Damage cause | Chemical composition | References |
|---|---|---|---|---|---|---|---|---|---|
| KJ-39, Krafla, Iceland | 2800, 9186 | 350/662 | Casing: 9 5/8" Slotted liner: 7 5/8" | Casing: Carbon steel, API 5L K55 Liner: Carbon steel, API 5L K55 | Cavitation damage and corrosion inside due to parts of the carbon steel liner corroding and breaking in the well that started only a few weeks after the opening of the fluid flow from the well | Wellbore was blocked at 1600 m corrosion, The liner was broken in half at 1600 m when trying to retrieve it | Uniform and pitting corrosion, hydrochloric acid and sulfide stress corrosion, cavitation corrosion, hydrogen embrittlement and cracking, thermal stresses | 4085 ppm $CO_2$, 560 ppm $H_2S$, 330 ppm HCL, 75 ppm $N_2$, 60 ppm $H_2$ gas in formation steam | (Thorbjornsson, 2012) |
| IDDP-1, Krafla, Iceland | 2102, 6896 | 450/858 | Casing: 9 5/8" Slotted liner: 9 5/8" | Casing: Carbon steel, K55, Hydril 563 Liner: Carbon steel, K55, BTC | Erosion caused by $SiO2$ precipitation, corrosion of production liner, nozzles in wellhead, hydrogen embrittlement of API K55 casing material. Master valve failure due to corrosion. API T95 were less affected by sulfide corrosion and hydrogen embrittlement. | Slotted liner, production casing, wellhead equipment | Hydrochloric acid and sulfide stress corrosion cracking, pitting corrosion, horizontal cracks and fissures parallel to the surface, hydrogen embrittlement, aggressive silicate precipitation, thermal stresses | 732 ppm $H_2S$, 93 ppm HCL, 10 ppm $H_2$ gas in formation steam | (Friðleifsson et al., 2015), (Markússon & Hauksson, 2015), (Hauksson et al., 2014) |
| IDDP-2, Reykjanes, Iceland | 4650/ 15256 | 427/ 800 | Casing: 9 5/8" Perforated liner: 7" Tubing: 3.5 in | Casing and perforated liner: Carbon steel, L80, BTC Tubing: Carbon steel, API 5DP PSL1 grade G-105 | Axial cracks in tool joints distributed evenly on the circumferential of the joint boxes, uniform and pitting corrosion in pipe bodies | A hole was created on casing as a result of corrosion at 2300 m, corrosion damage is observed in injection string from 4000 m to 4659 m that was inside a perforated liner | Sulfide stress corrosion cracking, thermal stresses, sulfide corrosion | 3.0E-07 mg/L $H_2S$ and 23.7 mg/L $CO_2$ gas in formation steam, 12 ppm $O_2$ in injection fluid | (Friðleifsson et al., 2017), (Karlsdóttir et al., 2019) |

Table 3. Summary of corrosion findings from KJ-39, IDDP-1 and IDDP-2 wells in Iceland

In the context of abrasive super-hot hard rock drilling, the selection of appropriate materials is crucial for mitigating corrosion and ensuring the longevity of drilling equipment. Table 3 provides a summary of the corrosion findings from KJ-39, IDDP-1, and IDDP-2 wells in Iceland, highlighting the severity of the corrosion problem in geothermal environments.

### 3.4. Innovative technologies for super-hot EGS drilling

The development of super-hot EGS holds immense potential for unlocking vast reserves of clean and sustainable geothermal energy. However, harnessing this potential requires significant innovation and advancements in technology. The extreme temperatures, challenging drilling conditions, and complex reservoir characteristics associated with super-hot EGS necessitate the development of specialized tools, materials, and techniques.

#### 3.4.1. Thermally Enhanced Drill Bits

To withstand the abrasive nature of hard rock formations at high temperatures, thermally enhanced drill bits have been developed with improved durability and wear resistance. These drill bits incorporate advanced materials and coatings, such as polycrystalline diamond compact (PDC) cutters and high-temperature alloys, to ensure efficient drilling performance and extended tool life. In previous super-hot EGS drilling projects, both roller cone and PDC bits have been tested and thermally enhanced PDC bits have proved to perform better. One of the primary limitations imposed by high temperatures in super-hot geothermal drilling is the challenge associated with the use of elastomers and temperature-resistant grease, particularly in sealing and lubricating the bearings of roller cone bits. However, innovative approaches have been explored to overcome this limitation. For instance, in the Venelle-2 well drilling within the DESCRAMBLE project in Italy, a special type of polycrystalline diamond compact (PDC) bit was employed (Bertani et al., 2018). Additionally, in the IDDP-2 well of the DEEPEGS project in Iceland, an elastomer-free tricone and hybrid bit, along with a specially designed high-temperature grease rated for up to 300°C (572°F), were utilized (Friðleifsson et al., 2017). These successful demonstrations of drill bit development highlight the progress made in addressing the limitations posed by temperature on drill bit performance. In addition to the choice of drill bit and specialized materials, other factors such as rotational speed (RPM), weight on the bit (WOB), drilling fluid flow rate and pressure, and the strength of the rock formation being drilled are crucial considerations in achieving optimal drilling performance and rate of penetration (ROP) (Nygaard & Hareland, 2007). The ROP is influenced by a combination of these factors, and understanding the specific properties of the rock formation is essential for selecting the appropriate drill bit type and optimizing drilling operations.

National Oilwell Varco (NOV) has developed the Phoenix Series Drill Bits, that integrate ION™ cutters, which harness NOV's patented thermal-stabilizing, deep-leach technology. The optimized cutter geometries strike an optimal balance between fracturing and shearing rock-failure mechanisms, thereby maximizing the efficacy of rock failure and facilitating superior drilling performance. Notably, the ION-shaped cutters have proven to be highly effective and efficient in volcanic rock formations, delivering increased drilling efficiency without compromising durability. Demonstrated successes include achieving an exceptional 67% higher rate of penetration with a single bit run in New Zealand, as well as notable gains in drilling distances up to 8% farther and drilling speeds up to 36% faster in geothermal operations conducted in Indonesia.

### 3.4.2. Directional Drilling and Steering Tools

Directional drilling poses significant challenges in super-hot drilling environments, primarily due to the sensitivity and fragility of sensor components and the incompatibility of elastomers with high surrounding temperatures. The extreme heat encountered in super-hot wells can lead to the deterioration and damage of sensors, which are crucial for precise wellbore placement and navigation. Additionally, the elastomers commonly used in directional drilling tools are not designed to withstand the exceptionally high temperatures, resulting in degradation and eventual failure. The reliance on accurate directional drilling and steering tools is crucial for controlling the well profile in super-hot geothermal systems.

Traditionally, conventional methods such as the pendulum BHA assembly have been employed as an alternative in situations where directional drilling tools face temperature limitations. However, these approaches often have limitations in terms of accuracy and control. Recognizing the need for advanced solutions, researchers have undertaken efforts to develop high-temperature directional drilling systems. For instance, (Chatterjee et al., 2015) presented a government-funded study on the development and testing of a 300°C (572°F) elastomer-free directional drilling system. The system was subjected to two field tests in the BETA field, drilling into granite rock below a measured depth of 4938 ft. The tests demonstrated promising results, achieving 15-20 feet/hour rate of penetration and 6°/100 ft steering with commercially viable efficiency, providing valuable insights into the potential of such technologies. Nevertheless, as super-hot geothermal systems aim to operate at temperatures around 500°C (932°F), there is a pressing need for further research and development to advance directional drilling and steering tools capable of withstanding and operating under such extreme conditions.

### 3.4.3. Drilling Fluids and Loss Circulation Materials

Drilling fluids play a vital role in maintaining pressure control during geothermal drilling operations, but the presence of supercritical temperatures introduces a significant challenge known as thermal degradation. Researchers (Vivas & Salehi, 2021) and (Mohamed, Salehi, Ahmed, et al., 2021) conducted laboratory experiments at temperatures up to 190°C to evaluate the thermal stability and effectiveness of different lost circulation materials (thermoset shape memory polymer (SMP)). Their studies revealed a considerable reduction in viscosity with increasing temperature due to thermal degradation and thinning of the drilling fluids. They concluded that temperature has a significant impact on the rheological properties, particularly the viscosity and gel strength of drilling fluids containing various chemical additives.

During drilling of Venelle-02 well (Bertani et al., 2018), Water Based Mud weighted up with Ilmenite (Microdense™) and with Sepiolite as suspending agent was used. The performance of the drilling fluid was significantly improved in terms of temperature resistance. Even after an extended period, the fluid exhibited no sagging and effectively controlled fluid loss. In the study, Ilmenite (Microdense™) was tested as the weighting agent. This weighting agent demonstrated exceptional properties due to its unique particle size distribution, with an average size of 5 μm. These characteristics provided auto-suspending properties, effectively preventing sagging and settling of the weighting agent within the fluid.

Loss circulation can result in substantial costs, accounting for more than 20% of the total exploration well drilling expenses (Lavrov, 2016). (Cole et al., 2017) investigated geothermal wells drilled between 2009 and 2017 and identified natural fractures as the primary cause of lost circulation. Despite attempts to mitigate the issue using various LCMs or cementing to seal the zones, major super-hot geothermal wells, as reported by (Batini et al., 1983; Bertani et al., 2018; Friðleifsson et al., 2017; Friðleifsson et al., 2015), have experienced partial to complete loss of circulation. Despite the availability of multiple LCM products and ongoing research projects to innovate prevention methods, the risk of loss circulation remains a significant concern for future geothermal drilling projects (Magzoub et al., 2021).

### 3.4.4. Materials for High-Temperature and Corrosive Environments

The selection of appropriate materials is crucial for mitigating corrosion and ensuring the longevity of drilling equipment. Table 2 below presents a summary of 113 days of corrosion tests using different materials in corrosive steam produced from IDDP-1 geothermal well (Karlsdóttir et al., 2015; Thorbjornsson et al., 2015). The materials tested are carbon steel, stainless-steel, as well as rarely used titanium and Ni-base alloys. The chemical composition of steam consisted of mainly $CO_2$ (339 mg/kg) and $H_2S$ (732 mg/kg) with average pH value of 2.7 at 240-270°C (464-518°F). Carbon steel, commonly used in manufacturing of pipes, drill string components and casings, exhibited the highest corrosion rate and extensive pitting corrosion. Austenitic stainless steels, such as the 304/316 types, showed limited resistance to corrosion, while higher alloyed austenitic steels demonstrated better performance and could be considered for use above 250°C. Duplex stainless steels exhibited excellent corrosion resistance below 250°C but should be avoided above this temperature due to structural changes and erosion-corrosion damage. Ni-base alloys, although not widely used in geothermal wells, showed evidence of corrosion pitting in N08825 type and small corrosion damage in the form of narrow pitting in N06625 type. Titanium alloys, specifically R50400 type, exhibited narrow pitting, while R52400 type demonstrated greater resistance. However, the use of titanium alloys at higher temperatures (>400°C) is limited due to strength concerns, and hydrogen embrittlement is a concern above 80°C.

### 3.4.5. Cementing and Well Integrity

Silica-modified Portland-based cement formulations, which are often used in oil and gas wells for high-temperature oil wells, are not resilient in harsh geothermal settings and do not effectively offer zonal isolation or metal casing corrosion-protection (Kang et al., 2022; Petty (a) et al., 2020; Shah, 2021). Furthermore, rapid temperature change during flow test and subsequent well killing operations and temperature difference between drilling fluid and formation drilled result in casing buckling (Chiotis & Vrellis, 1995). According to (Bertani et al., 2018), a thermally stable cement, ThermaLock™ developed by Halliburton was used in Descramble project in all the cementing jobs of 7" casing and liner and for the temporary plug and abandon job without encountering poor cementing quality problems. ThermaLock™ is a non-portland calcium phosphate cement system with superior thermal stability above 230°F (110°C) and it can be augmented with mechanical property enhancers for extreme environments with significant thermal and stress cycles. (Halliburton, Accessed 28 Jun. 2023). Figure 2 below shows visual effect of $CO_2$ deterioration of Portland cement over time, while leaving ThermaLock cement virtually unaffected. Schlumberger developed ThermaSTONE thermally responsive cement system expands and contracts with the high pressures and temperatures encountered in geothermal environments. According to (Tomilina & Chougnet-Sirapian, 2012), lab testing at 343°C (650°F) showed that the ThermaSTONE system retained tensile strength for more than 6 months whereas the tensile strength of conventional cementing systems decreased over time.

| Material | | Chemical composition of steam, (mg/kg) | Corrosion rate (mm/year) | Localized corrosion | | Stress-Corrosion Cracking (SCC) | | Notes |
|---|---|---|---|---|---|---|---|---|
| | | | | Pits | Cracks | 20% true | 12.5% true | |
| Carbon steels | G10150 | $CO_2$ - 339<br>$H_2S$ - 732<br>$H_2$ - 10<br>Cl - 93<br>F - 5.0<br>Fe - 8.4<br>B - 1.0<br>pH - 2.7 at 240–270°C | 0.0064 | + | – | - | - | Most common material used in geothermal wells. Corrosion rate was the highest. Pitting corrosion was quite extensive for the TN95. |
| | K55 | | 0.0051 | + | – | N/A | N/A | |
| | TN95 | | 0.0058 | + | – | N/A | N/A | |
| Austenitic stainless steels | S30403 | | 0.0017 | + | + | + | + | Widely used in surface power production equipment. The 304/316 types had limited resistance to corrosion. The higher alloyed austenitic steels had better performance. Good candidates for use above 250°C. |
| | N08904 | | 0.002 | + | + | + | + | |
| | S31603 | | 0.001 | + | - | - | - | |
| | N08028 | | 0.0011 | + | - | - | - | |
| | S31254 | | <0.0003 | + | + | N/A | N/A | |
| Duplex stainless steels | S32750 | | 0.0006 | + | - | N/A | N/A | Excellent resistance to corrosion below 250°C. Not recommended to use above 250°C due to its structural change. Erosion-corrosion damage observed in S32707. |
| | S32707 | | 0.0006 | + | - | N/A | N/A | |
| Ni-base alloys | N06255 | | <0.0003 | + | - | N/A | N/A | Not been widely used in geothermal wells. N08825 type suffered from corrosion pitting. N06625 type showed evidence of small corrosion damage in the form of narrow pitting. |
| | N08825 | | 0.0034 | + | + | N/A | N/A | |
| | N06625 | | 0.0007 | + | + | N/A | N/A | |
| Titanium alloys | R50400 | | N/A | + | + | N/A | N/A | R50400 type suffered from narrow pitting. R52400 type showed more resistance. The concern of strength limits the use at higher temperatures (>400°C). Hydrogen embrittlement is also concern above 80°C. |
| | R52400 | | N/A | + | - | N/A | N/A | |

**Table 3. Summary of 113 days of corrosion tests using different materials in corrosive steam produced from IDDP-1 geothermal well in Iceland. (Karlsdóttir et al., 2015; Thorbjornsson et al., 2015)**

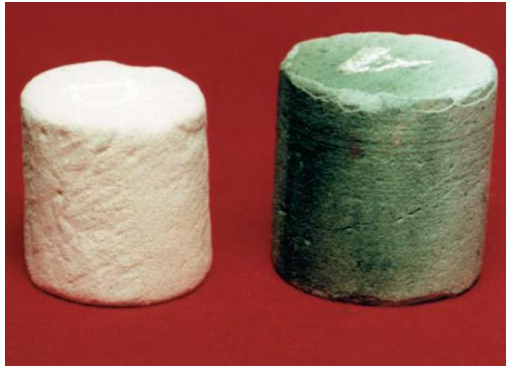

**Figure 2. A visual comparison of ThermaLock cement (right) and Portland cement (left) under CO2 exposure.**

### 4. Techno-Economic Analysis

#### *4.1 Methodology for analysis*

The Techno-Economic Analysis section of this paper aims to assess the feasibility and economic viability of geothermal systems operating within the temperature range of 170°C to 370°C (338°F to 698°F). By evaluating the costs, potential returns on investment, and financial aspects associated with the development and operation of such systems, this analysis provides valuable insights into their techno-economic feasibility (Khankishiyev & Salehi, 2023). The Geothermal Electricity Technology Evaluation Model (GETEM) (Mines, 2016) is used to estimate the Levelized Cost of Electricity (LCOE) and well cost for multiple downhole temperatures. GETEM is an excel-based detailed model of the estimated performance and costs of currently available U.S. geothermal power systems.

The measured depth input is determined to be 4 km (13123 ft) while deeper wells can be simulated using this tool. According to underground temperature maps at 4.5 km by (Blackwell et al., 2011), temperatures from 150°C to 350°C (302°F to 662°F) can be encountered across United States, while the some super-hot EGS well drilling activities around the world showed that super-critical temperatures (400-500°C (752-932°F) are possible to be untapped (Batini et al., 1983; Bertani et al., 2018; Friðleifsson et al., 2017). However, the GETEM tool is limited to 373°C (703.4°F) and it cannot run the calculations above critical temperature of the water. Therefore, the temperature range used in calculations was decided to be from 150°C to 370°C (302°F to 698°F).

Rate of penetration (ROP) is one of the most important factors determining the total well cost that necessitates its optimization. Previous projects showed that (CATF, 2022; Kruszewski & Wittig, 2018) drilling through hard rock formations at elevated temperatures are very challenging, limiting ROP between 120 ft/day to 400 ft/day. For the purpose of techno-economic evaluation, ROP values of 120-1200 ft/day were used in GETEM to compare total well costs and LCOE. The model considers both flash-steam and binary power plants for geothermal power generation. However, it was recommended that Binary system to be used above 200°C (392°F) and Flash system not to be used below 150°C (302°F).

#### *4.2. Cost estimation and economic viability*

Figure 3 below illustrates the change in total well cost in million US$ with increasing rate-of-penetration in ft/day for the well with measured depth of 4 km (13123 ft). Increasing to ROP from

150 ft/day to 500 ft/day can decrease the well cost by twice. The percentage distribution of the cost categories considered by the GETEM has been given in Figure 4 for the well with MD of 13123 ft drilled with 270 ft/day rate (GETEM default). The rig, cementing, casing and directional drilling are the top 4 categories consisting of 3/4 of the total well cost. Although the mud cost is only at 5%, it can go up significantly during partial or total fluid loss events during drilling due to widely distributed fracture networks.

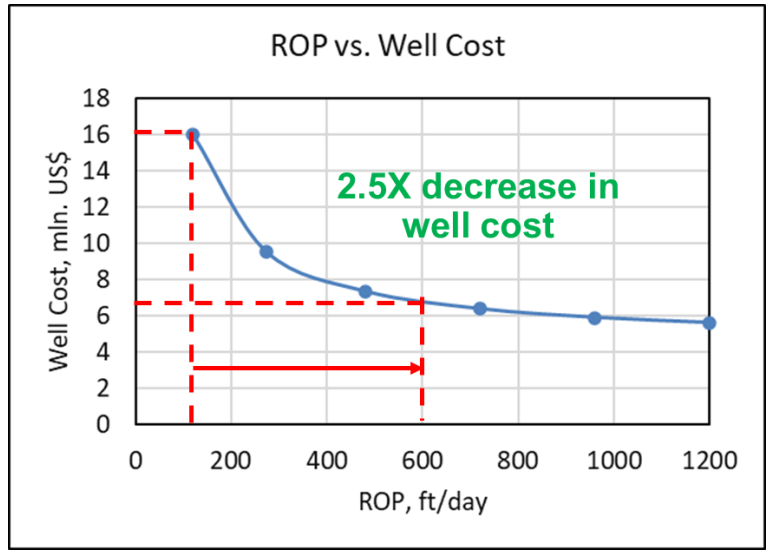

**Figure 3. Well cost (mln. US$) vs Rate-of-penetration (ft/day)**

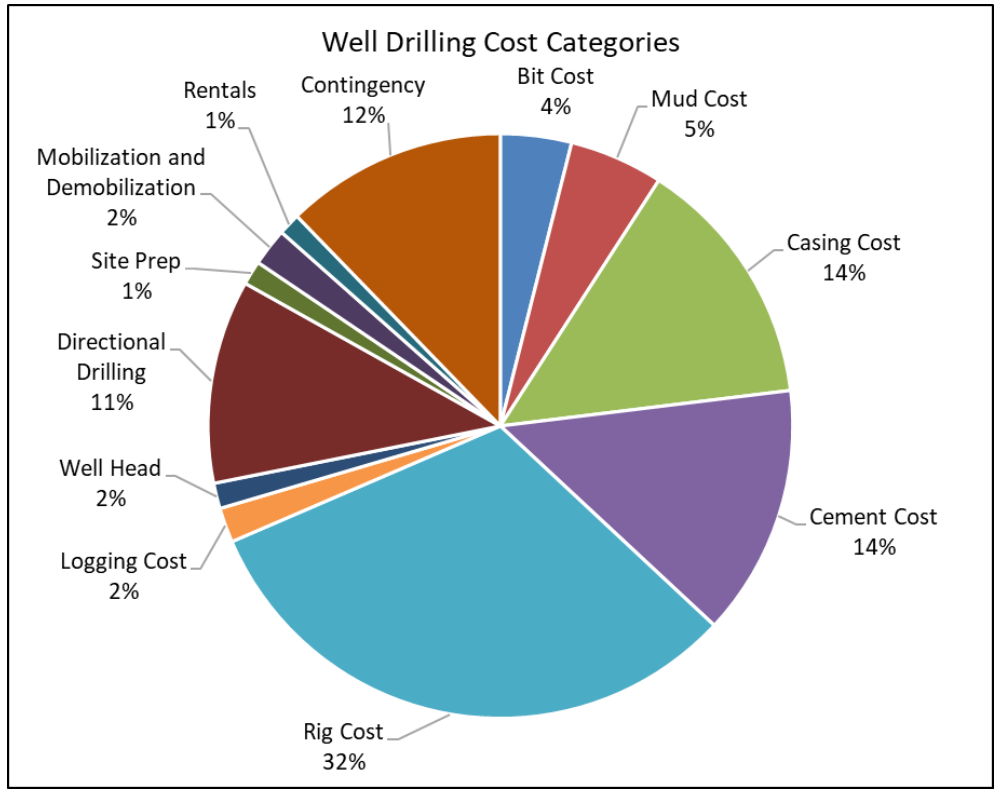

**Figure 4. Geothermal well drilling cost categories**

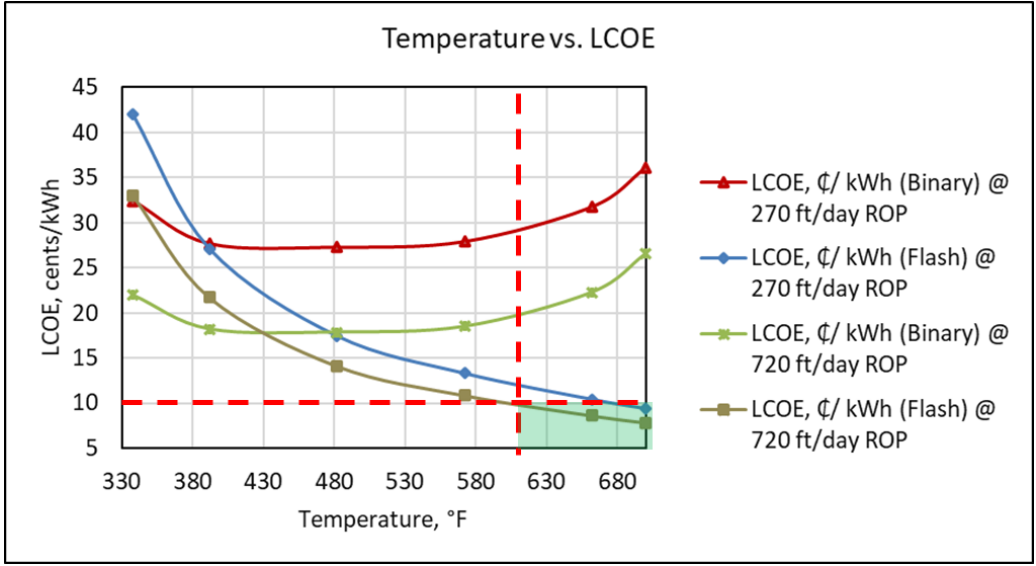

**Figure 5. Temperature (°F) vs. LCOE (cents/kWh)**

Figure 5 above illustrates the impact of resource temperature on the levelized cost of electricity (LCOE) for both flash and binary geothermal systems, spanning a range from 340°F to 700°F (170°C to 370°C). It is evident that beyond approximately 400°F (200°C), the LCOE for binary systems starts to increase, while below this threshold, the LCOE for flash steam systems exceeds that of binary systems. In contrast, Figure 6 below showcases the influence of the rate of penetration (ROP) and total well cost on the levelized cost of electricity for a resource temperature of 700°F (370°C). The combined insights from these two figures lead to the conclusion that resource temperature has a substantial impact on reducing the LCOE than the ROP. Therefore, it becomes imperative to pursue higher resource temperatures to enhance the return on investment. However, it is essential to consider that achieving higher temperatures may necessitate drilling significantly deeper wells, resulting in a potential exponential increase in well costs.

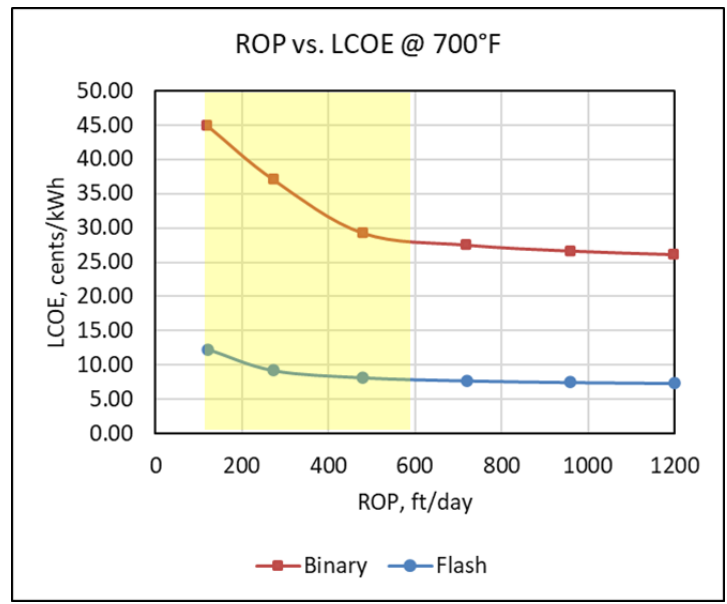

**Figure 6. ROP (ft/day) vs. LCOE (cents/kWh)**

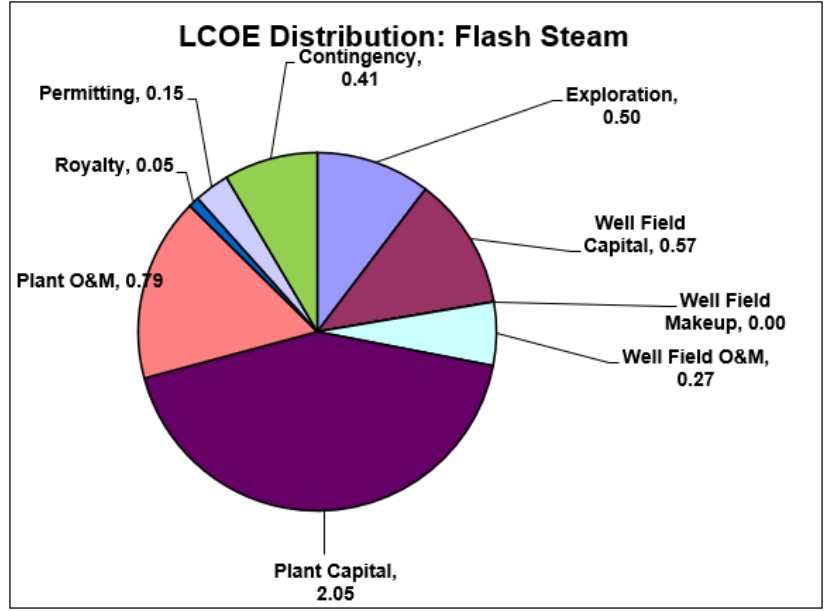

**Figure 7. LCOE distribution of the flash steam power plant at 700°F (370°C)**

According to Figure 7, the power plant capital cost constitutes the largest portion of the levelized cost of electricity (LCOE) for geothermal power. This highlights the significance of efficient investment in power plant infrastructure. Additionally, the development and exploration of the well field, including the optimization of producer and injector wells, plays a crucial role in achieving economic efficiency. To enhance the economic performance of geothermal power generation, it is important to strategically locate power plants near abundant high-temperature resources that are in proximity to the electrical grid. This proximity reduces transmission losses and enhances the overall economic viability of the project. By optimizing the number and placement of wells in the geothermal field, the production capacity can be maximized, leading to improved economic performance and a lower LCOE.

**Conclusion**

In conclusion, this study has explored the potential of super-hot enhanced geothermal systems (EGS) in hard rock formations and highlighted the significant advancements and challenges in this field. The analysis of conventional geothermal systems has revealed their limitations, particularly in handling the extreme temperatures associated with super-hot EGS. However, through innovative technologies such as advanced drilling techniques, specialized drill bits, and high-temperature materials, significant progress has been made in addressing these limitations. The successful development and application of tools like the Phoenix Series Drill Bits and elastomer-free directional drilling systems demonstrate the effectiveness of these technological advancements.

The thermal degradation of drilling fluids and the issue of loss circulation have been identified as critical challenges in super-hot drilling. Laboratory experiments have provided valuable insights into the thermal stability and rheologic properties of drilling fluids, facilitating the development of temperature-resistant additives and mitigation strategies. The study of corrosion resistance in different materials has revealed the varying performance of carbon steel, austenitic stainless steels, duplex stainless steels, Ni-base alloys, and titanium alloys in corrosive geothermal environments.

These findings underscore the importance of carefully selecting materials that offer the necessary corrosion resistance and structural integrity for long-term geothermal operations.

From a financial standpoint, the selection of materials must consider both the initial costs and the long-term cost-effectiveness. Investing in high-performance materials that can withstand the extreme conditions of super-hot EGS can lead to reduced maintenance and replacement costs over the project's lifespan. Furthermore, the economic viability and potential returns on investment must be evaluated through techno-economic analysis, considering factors such as project scale, resource potential, and market conditions.

All in all, the development of super-hot enhanced geothermal systems in hard rock formations presents immense potential for sustainable and reliable energy production. The advancements in drilling technologies, materials, and fluid management discussed in this paper have paved the way for future research and development in the field of geothermal energy. However, further research is still needed to address remaining challenges, such as directional drilling, well integrity, and environmental considerations. By continuing to innovate and refine these technologies, the geothermal industry can unlock the full potential of super-hot EGS and contribute to a cleaner and more sustainable energy future.

## Acknowledgement

The authors of this paper would like to thank DeepPower Inc for funding this research. The opinions, findings, conclusions, or recommendations presented in this publication are solely those of the authors and do not necessarily represent the views or opinions of DeepPower Inc.